# An Automated Evaluation Metric for Chinese Text Entry

**ABSTRACT**

In this paper, we propose an automated evaluation metric for text entry. We also consider possible improvements to existing text entry evaluation metrics, such as the minimum string distance error rate, keystrokes per character, cost per correction, and a unified approach proposed by MacKenzie, so they can accommodate the special characteristics of Chinese text. Current methods lack an integrated concern about both typing speed and accuracy for Chinese text entry evaluation. Our goal is to remove the bias that arises due to human factors. First, we propose a new metric, called the correction penalty (P), based on Fitts' law and Hick's law. Next, we transform it into the approximate amortized cost (AAC) of information theory. An analysis of the AAC of Chinese text input methods with different context lengths is also presented.

**Author Keywords**

Text entry, automated evaluation, context length, error rates.

**ACM Classification Keywords**

H5.2. Information interfaces and presentation (e.g., HCI): User Interfaces.

**INTRODUCTION**

Text entry on mobile devices has been well-studied in the field of human-computer interaction (HCI). For example, text input methods, such as MultiTap, T9, and LetterWise, have been evaluated by KSPC (Keystroke Per Character) [7], MSD (Minimal String Distance) [7], unified error metrics [8], and CPC (Cost Per Correction) [1]. All of these metrics focus on alphabetic languages, mostly English. However, thus far, research on text entry in other languages has been under exploited.

Since most ideograph-based Asian languages consist of thousands of complex characters, it would be impractical to create a huge keyboard to map every possible character. Modern GUI environments, whether Microsoft Windows, Mac OS X, or X11, come with built-in tools for transforming multiple composition keystrokes into a single ideograph. These tools, known as "input methods", are often categorized into "radical-based" or "phonetics-based" methods. With radical-based input methods, users generate a character by typing the composing radicals, whereas with phonetic-based approaches, users generate a character by typing the syllables. In both methods, if there are homo-radicals and homo-phonics, respectively, a choice has to be made and the proper character has to be selected and entered.

Due to the complicated input process of Asian languages, localized versions of mobile phone text entry methods present unique challenges. Chinese T9, for example, behaves more like a desktop input method than its English sibling. As Figure 1 shows, Chinese T9 Pinyin involves prediction in two stages: the first stage lists proper syllables and their corresponding Chinese characters, while the second suggests candidates for the next Chinese characters. Thus far, only a few works have addressed the human factors in Asian language text input methods [4, 9]. Most Asian text input methods focus more on a system's performance and evaluate its accuracy in terms of character error rate (CER) only [2, 10].

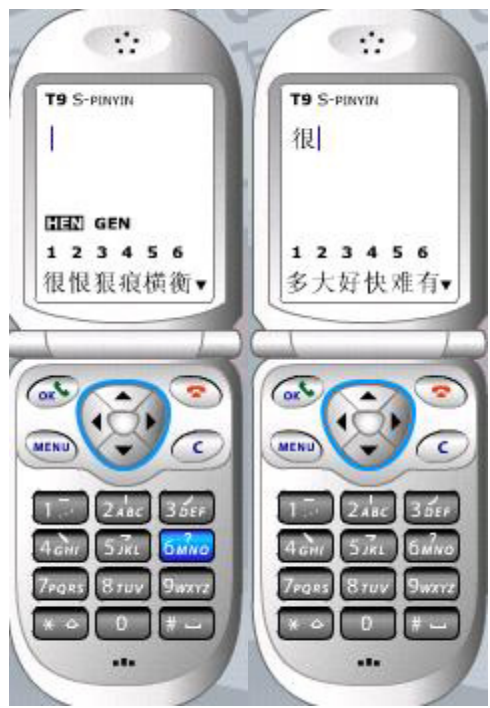

**Figure 1. Two stages of prediction in Chinese T9 Pinyin method[1]. Stage 1: Key pressed order: 4->3->6. In this case, T9 predicts two valid combinations of Pinyin and lists the most frequent characters generated from them. Stage 2: After users choose a word, T9 suggests candidates for the next Chinese character.**

In this paper, we propose an automated approach with a new metric, which is related to CER and CPC, to evaluate the efficiency of Chinese input methods for desktop environments. Our objective is to connect research in Chinese text input methods to the HCI area and thereby gain some insights for improving the usability of text entry.

---

[1] Originally from www.t9.com

## RELATED WORK

### Fitts' Law

Fitts' law was first used in HCI by Card et al in 1978. It is a function of the distance to the final target and the size of the target, that is used to predict the time required to move rapidly from a starting position to a final target area. Mathematically, Fitts' law can be formulated in several ways. One refined form, proposed by MacKenzie [8] is:

$$T = a + b\log_2(D/W + 1),$$

where T is the average time taken to complete the movement; and a and b are empirical constants that can be determined by fitting a straight line to measured data. D is the distance from the starting point to the center of the target. W is the width of the target measured along the axis of motion. The term $\log_2(D/W + 1)$ represents the index of difficulty (ID) of the given task. Since text entry task usually shifts the cursor by keystrokes, rather than mouse movements, ID may link to the number of keystrokes directly.

### Hick's Law

When correcting typing errors, both the time taken by the ID of the task and the time for candidate selection should be considered. Hick's law,

$$T = b\log(n + 1),$$

describes the time, T, it takes users to make a decision as a function of the equal possible *n* choices they have, where b is an empirical constant. The law gives us some baseline points, but the realistic candidate selection time still needs to be measured via subject experiments [8]. As far as we know, Hick's law has not been widely adapted to candidate selection for the correction typing error in text entry tasks.

### CER, Language Model and Perplexity

CER is the most widely adopted metric for performance evaluation of desktop text input methods, especially in Asian languages. To predict the baseline performance of a specific language, the n-gram language model and its perplexity are useful. Perplexity can be thought as the weighted average number of choices for the "next word." It is expected that fewer choices reduces the CER. The language model is built from large amounts of text data, which is left after the text input task has been completed. According to MacKenzie [6], the editing processes have to be captured into the input-based language models for text entry research.

### MSD, KSPC, and Unified Error Metrics

Real-time typing error correction of computer text entry makes the recognition of error modification complicated. Errors in computer text entry can be arranged into two classes: those not corrected in transcribed text, and those that are corrected. The latter do not appear in the transcribed text and have been often ignored in previous works. In this section, we describe well-known evaluation metrics for text entry and consider their shortcomings.

### MSD

Evaluating the accuracy of text input entry involves more than simply comparing strings. Consider the following example:

Presented Text:   the quick brown fox

Transcribed text: the qui**xck br**wn fox

In the above, there are six errors in the string comparison. However, most people would delete "x" and then add "o" instead of modifying all six errors. The notion of MSD (minimum string distance) is introduced to deal with such a situation [7]. In brief, the minimum string distance between two strings is the minimum number of primitives – insertions, deletions, or substitutions – needed to transfer one string to another. The MSD in this case is 2. The original MSD error rate was defined as follows:

$$Old\ MSD\ Error\ Rate = \frac{MSD(P,T)}{\max(|P|,|T|)} \times 100\%,$$

where | | shows the length of the text data. The MSD error rate may be unreliable when insertions and deletions both occur. Thus, we need to consider the alignments between presented text and transcribed text. The refined definition of MSD is as follows:

$$New\ MSD\ Error\ Rate = \frac{MSD(P,T)}{\overline{S_A}} \times 100\%,$$

where $\overline{S_A}$ is the mean length of the alignment strings.

By comparing presented text and transcribed text, MSD can only provide information about the remaining text because error corrected by the editing process can no longer be observed.

### KSPC

In contrast to MSD, it is possible to observe corrected errors under KSPC by logging all keystrokes as an input stream. From the input stream, a metric, KSPC (Key Strokes Per Character), is defined [8].

KPSC = |InputStream| / |TranscribedText|.

Different from the MSD error rate, KSPC only describes the effort required to correct errors without considering uncorrected errors. A large number of errors that only require low correction effort and a few errors requiring high correction effort may result in the same KSPC value. The keystrokes that create errors and keystrokes that correct errors are different, but they are not distinguished by KSPC.

**MacKenzie's Unified Error Metrics**
After observing the shortcomings of the MSD error rate and the KSPC value, MacKenzie et al [8] proposed a unified error metric that logs the input stream in the same way as KSPC and then classifies the keystrokes to analyze the transcribed text, According to MacKenzie, the keystrokes can be divided into four categories:

- C: Correct. This class contains all keystrokes that are correct characters in the transcribed text.
- INF: Incorrect Not Fixed. The keystrokes resulting in unexpected characters remaining in the transcribed text are in this class.
- IF: Incorrect Fixed. Keystrokes in this class generate errors that are corrected later.
- F: Fixes. Editing function.

In brief, only keystrokes in F are not represented by any character. Keystrokes in C and INF comprise characters in the transcribed text, while errors corrected correspond to keystrokes in IF. The MSD is only concerned with INF, while KSPC only reports the sum of IF and F. A metric that can identify all keystrokes in INF and IF is defined as follows:

$$Total\ Error\ Rate = \frac{INF + IF}{C + INF + IF} \times 100\%.$$

MSD error rate and KSPC statistic can be defined in terms of the keystroke taxonomy,

$$MSD\ Error\ Rate = \frac{INF}{C + INF} \times 100\%,\ and$$

$$KSPC \approx \frac{C + INF + IF + F}{C + INF}.$$

**Cost Per Correction**
Recently a new metric called CPC (Cost per Correction) [1] was proposed for calculating the effort a user makes by having to use extra keystrokes to handle corrections. It calculates the effort as follows,

$$CPC = \frac{WLS + FS + WCS}{NoC},$$

where WLS denotes the number of wasted letter key strokes, FS represents the number of keystrokes used for corrections, WCS denotes the number of wasted control key strokes, and NoC indicates the number of corrections. Intuitively, a lower value of CPC should generate higher user satisfaction. However, overemphasis on a lower CPC does not necessarily lead to a greater satisfaction since there may be too many corrections.

According to the metrics discussed above, it is therefore necessary to consider an alternative approach to overcome the shortcomings of existing metrics.

### THE TWO CULTURES OF TEXT ENTRY
Character-based and non-character-based languages are intrinsically different. One possible major difference is the detection of word boundaries. For example, in English, words are naturally segmented by spaces, while there are no explicit boundaries in Chinese. To explain further, we describe the text entry processes in both languages below.

First, we consider character-based languages. Clearly, all words in character-based languages are composed of dozens of letters. Their input methods can be divided into two phases:

(1) Users type keystrokes to generate letters

(2) Input methods collect letters to produce meaningful chunks, e.g, words.

In the first phase, it is relatively simple to design a mapping from keystrokes, or compositions of keystrokes to letters, since the alphabet of most languages contains less than 40 letters. In recent years techniques, such as word prediction, have been widely used to improve the performance of the second phase by providing a candidate list based on the word stem typed in the first phase.

In contrast to character-based languages, there may be thousands of independent characters in non-character-based languages. A practical input method would define a set of components, be it phonetics or radicals, and then encode each independent character as a composition of components in the set. For example, Chinese character "天" (sky) is encoded as phonemes "ㄊㄧㄢ" in Phonetic, radicals "一大" in CangJie[2], and phonemes "tian" in Pinyin. The process can be decomposed as follows:

(1) Users type keystrokes to generate components

(2.1) Input methods transfer components to independent characters.

(2.2) Input methods collate characters to produce words

The first phase is similar to that of character-based languages. The input methods manage the second phase. In radical-based methods, the perplexity is very low. Most compositions of radicals correspond to unique character. Phonetic-based methods are limited by natural phonetic rules. In fact, there are about four hundred phonetics in Mandarin for thousands of Chinese characters. "yi" in

---
[2] CangJie is a system by which Chinese characters may be entered into a computer by means of a standard keyboard.

Pinyin maps to "一"(one), "衣" (dress), "醫" (cure), or even more. Thus, the perplexity is relatively high. Due to the diversity of the culture in text entry, the development and evaluation of Chinese text entry is widely different.

**MOTIVATION**
Many works have evaluated text entry performance using the results of participants' typing experiments. Though this approach may be practical, it has certain limitations. For example, human typing experiments are time-consuming, and difficult to verify or duplicate. In addition, many human factors can affect the results: degree of expertise, level of concentration, and so on. Finally, the works focus primarily on alphabetic languages. To evaluate more contexts in various devices and languages, some automatic experiment strategies for text entry with tolerable level of inaccuracy are thus required.

Intelligent Chinese input methods select a character automatically from among several candidates composed of the same components. Although the technique is similar to English T9, which predicts words with a dictionary-based model, blurred boundaries in Chinese make predictions complicated. Thus, intelligent input methods need to consider context lengths which affect the prediction result. Words in the input buffer affect the result of automatic homo-phonic text candidate selection; however, they may be modified after the next character is entered. By relying on the context length effect in intelligent input methods, most users tend to continue typing until the sentence is finished, even if they are aware that errors have occurred during the editing process,. In other words, they expect the input method to fix the error automatically; otherwise, they have to move the cursor to correct them.

Theoretically, a long context length, which carries more information content, should improve the accuracy of the input method's prediction. However, the error occurs in the early part of the input buffer, it may cost too much to correct. With the proper context length, intelligent input methods can achieve a trade-off between the error rate and the correction penalty. To the best of our knowledge, the unique characteristics of Chinese text entry have not been addressed previously.

**APPROXIMATE AMORTIZED COST OF CHINESE TEXT ENTRY**
Applying the information theoretic point of view to text entry, the amortized cost (A) of text input can be modeled using MacKenzie's unified metrics, as follows:

$$A = \frac{WastedBandwidth}{UtilisedBandwidth}$$

$$= (\frac{INF + IF + F}{C + INF + IF + F}) / (\frac{C}{C + INF + IF + F})$$

$$= \frac{INF + IF + F}{C}.$$

in which the basic unit of measurement of the four categories is the number of characters. Even though some might argue that the F, not characters, should be counted as keystrokes, we can successfully map each keystroke F to a specific unseen character as an information term. As long as the same unit is used to measure the numerators and denominators, the definition is satisfied.

| Situation | Input Progress | INF | IF | F |
|---|---|---|---|---|
| $S_0$ | No character fixed | $INF_0$ | 0 | 0 |
| $S_i$ | Some characters fixed | $INF_i$ | $IF_i$ | $F_i$ |
| $S_{all}$ | All character fixed | 0 | $IF_{all}$ | $F_{all}$ |

**Table 1. Three situations of errors correction.**

Whenever text input evaluation involves corrections, human factors must be considered because we cannot predict how the user will correct the errors. For example, in Table 1 there are three situations, but only situation $S_0$ lends itself to automatic experiments because there is no concern about corrections.

In character-based text entry, if we assume that the same number of errors occur in different situations and users adopt the same method to fix the problems, we would still need to design a keystroke event logger to record all the editing processes in order to determine the boundary of the amortized cost:

*NumberOfIncorrectCharacters*
$= INF_0 + IF_0 = INF_i + IF_i = INF_{all} + IF_{all}.$

$\because IF_0 = 0, \quad INF_{all} = 0$

$\Rightarrow INF_0 = IF_{all}.$

$A_0 \leq A_i \leq A_{all}$

$\Leftrightarrow \frac{INF_0}{C} \leq A_i = \frac{INF_i + IF_i + F_i}{C} \leq \frac{IF_{all}}{C} + \frac{F_{all}}{C} = \frac{INF_0}{C} + \frac{F_{all}}{C}.$

Unlike the evaluation of character-based text input, it is difficult to apply automated evaluation techniques to Chinese text entry. Because Chinese character generation involves two phases, it is not sufficient to map F using the same unit of measurement as other categories, i.e., counting the characters one-by-one. In other words, if a backspace keystroke is logged, it is difficult to determine whether a Chinese or a phonetic character has been erased; thus, evaluating the cost is problematic.

For this reason, we define the metric average penalty (P) and an amortized modification cost (M) instead of the original term, $\frac{F_{all}}{C}$, as follows:

$$P = \frac{T_H * INF_0 + T_F * \max(D_w)}{C + INF_0},$$

$$R = \frac{C}{C + INF_0},$$

$$\Rightarrow M = \frac{P}{R} = \frac{T_H * INF_0 + T_F * \max(D_w)}{C},$$

where $D_w$ denotes the distance that the cursor must be moved to correct farthest wrong word; $T_H$ indicates the time taken to choose candidates according to the Hick's law; $T_F$ represents the time required to move the cursor over the distance $D_w$ based on Fitts' law.

Next, we propose the following approximate amortized cost (AAC) to assess the performance of a text input system based on automated evaluation:

$$AAC = \frac{INF_0}{C} + M$$

$$= \frac{INF_0}{C} + \frac{T_H * INF_0 + T_F * \max(D_w)}{C}.$$

**EXPERIMENTS**

Measuring the time spent on choosing candidate words ($T_H$) and moving the cursor to the error to be corrected ($T_F$) is a complicated process. Many situations can occur during candidate selection, such as users use a numeric key to make a choice, the correct selection appears on the next page of the candidate form, etc. The time also varies from person to person depending on how familiar they are with text entry and how long they take to make a decision. Clearly, it is impossible to quantify these two factors without having real-time experiments. In our experiment, we assume that all candidates are equally likely and the time taken to choose candidates is the same. When using the QWERTY keyboard, most users type with all their fingers. The method of evaluating $T_F$ in our experiment is different from that of evaluating $T_F$ on a mobile keypad because only the thumb is used the latter case. Thus, the value of $T_F$ is linked to the distance that the cursor is moved.

**Apparatus and Software**
Seven computers, both desktop PCs and notebooks, are used as input devices with Going 8 IME, Microsoft New Phonetic 2003 IME, Microsoft ChangJie 2003 IME, and Chewing IME on Windows XP with SP2.

**Experiment Design**
The raw data for the presented text is comprised of 30,000 sentences retrieved from the ASBC corpus [3]. The independent variable in our experiment is the context length. To simulate the true typing process, we preprocess the presented text before transforming it into related keystrokes. The process is simple: given some integer k as the context length, if the length of the presented text, T, is shorter than k, we do not change T. Otherwise, we chop the first k characters of T to form a new string, denoted as $T'$, and then preprocess $T'$ in the same way. For example, in the experiment with input buffer length 3, "ab" remains and "abcdefgh" generates three strings: "abc", "def", and finally "gh" in the preprocessing phase. Each Chinese character corresponds to a unique composition of radical components, but may decode to several compositions of phonemes. Consequently we can easily transform presented texts to radical formats, and then keystrokes. In contrast, phonetic-based input methods require one more step. We use SRILM[4] to train a language model to segment the presented text into as many as phrase as possible. We then transform the presented texts to phonetic formats by looking up a dictionary, and generating keystrokes for our experiments.

After the preprocessing steps, the input methods adopt an input buffer of maximum length k. The input devices installed are: Going 8, Microsoft New Phonetic 2003, Microsoft ChangJie 2003, and Chewing input methods. Since these devices have never been used, their online-learning features have not been enabled, which ensures the experiment is unbiased. Keystrokes corresponding to the raw data simulate the input events for the above IMEs. Text output generated via the IMEs is then recorded. The number of characters and the number of keystrokes of C and INF are counted by comparing presented text with the output text, and by looking up the corresponding keystrokes of the incorrect characters.

---

[3] Academia Sinica Balance Corpus

[4] SRILM is a toolkit for building and applying statistical language models (LMs), primarily for use in speech recognition, statistical tagging and segmentation.

## Result

### MSD Error Rate

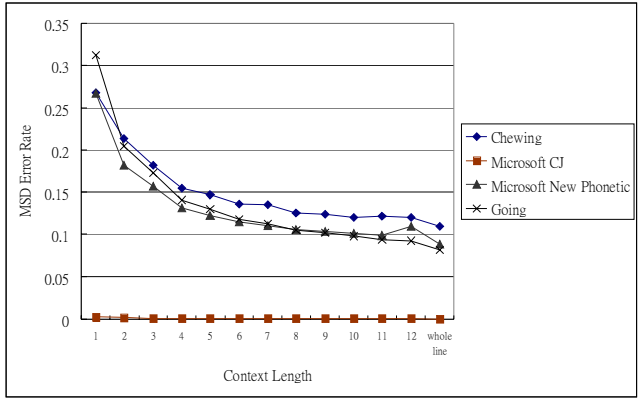

**Figure 2. Comparison of MSD Error Rate.**

### Correction Penalty

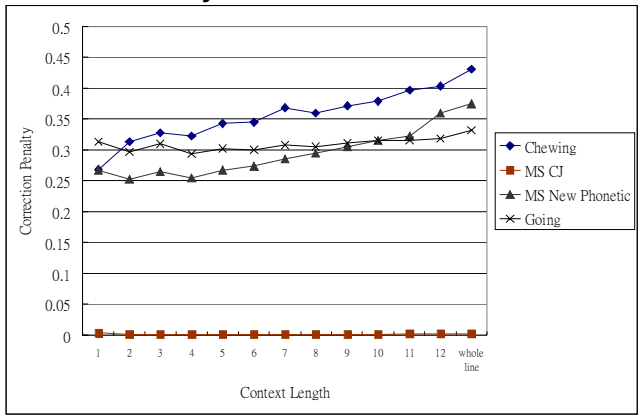

**Figuure 3. Comparison of Correction Penalty.**

### KSPC

| Phonetic-Based | Radical-Based |
|---|---|
| 3.15 | 4.48 |

**Table 2. KSPC values on QWERTY keyboard.**

### Approximate Amortized Cost (AAC)

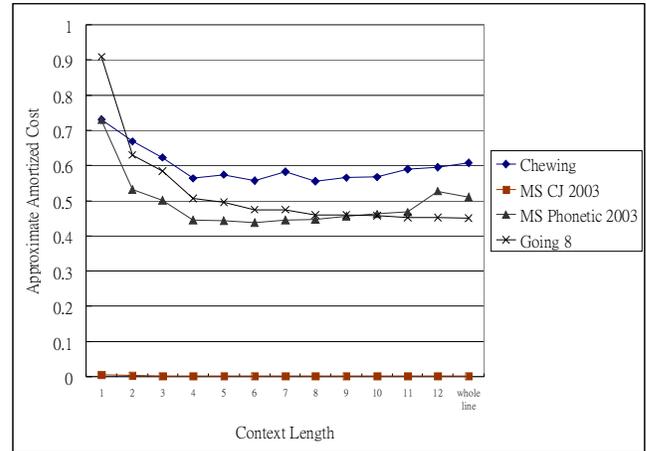

**Figure 4. Comparison of Approximate Amortized Cost.**

## DISCUSSION

We apply this automated experiment to four different IMEs. Microsoft New ChangJie IME is a radical-to-character conversion system. Chewing, Microsoft New Phonetic 2003, and Going 8 are phoneme-to-character conversion systems. Each of them is used to experiment on 13 kinds of different context length. Compared to the phonetic-based system, there are far fewer candidates for homo-radical characters. Thus, the MSD error rate of Microsoft New ChangJie 2003 approaches zero. Although based on the same phonetic input keystroke, the MSD error rate varies in the three different phoneme-to-character conversion systems. However, the MSD error rate of these three IMEs declines because of the longer input buffer string. In contrast, the correction penalty result shows the opposite trend (see Figure 3). The approximate amortized cost result (see Figure 4) shows the best balance between the correctness and the correction penalty. This helps us determine the context length for IME development.

For phonetic-based Chinese input methods, the best context length at the minimal AAC is around six-character long. This result suggests that it is possible to improve the usability of current automatic phoneme-to-character conversion system by fixing the window size of prediction to six. A system may consider picking proper transcribed result automatically within only six characters from current position to save a user's precious time of cursor movement and candidate selection without decreasing the correctness too much.

In the situation of Going 8, however, the window size could be larger, since its curve is smoother than MS Phonetic 2003 and Chewing. This difference may be related to their algorithms; MS Phonetic 2003 uses statistical language modeling approach [10], and Going 8 applies semantic pattern matching techniques [2]. Which means AAC may also provide another perspective besides MSD error rate to evaluate the design of automatic phoneme-to-character conversion systems.

**CONCLUSION AND FUTURE WORKS**

We have proposed a novel automated approach for text entry evaluation of Chinese language input methods. Because it is impossible to simulate every type of behavior in automated evaluation experiments, our goal is to minimize inaccuracy in real-time experiments. Further more, the modification process of Chinese text entry is quite different to that of character-based text entry (e.g. in English). Therefore, to combine the time taken by cursor movements and choosing candidates, and the amortized cost of information theory, we propose a metric, called the approximate amortized cost (AAC) to evaluate input methods of Chinese text entry.

We also compare the difference of several Chinese input methods using our proposed metric, approximate amortized cost (AAC). Because of the higher perplexity of the Chinese language model, it is also important to determine the accuracy of each input method in choosing suitable candidate words automatically. In addition, we calculate the *AAC* of each input method with different context length. The result helps us choose the appropriate context length at the minimum cost for each method in real-time.

In our future work, we will conduct real-time typing experiments to support our automated approach, which will allow us to approximate the real cost of text entry. We will also apply the evaluation approach to different input systems to reflect the trade-off between speed and accuracy. To build a flexible evaluation environment, we will apply OpenVanilla [3], an open source text entry and processing architecture, to conduct real text entry applications. With OpenVanilla's ability of supporting several desktop operating systems such as Windows, Mac OS X and Linux, the text entry applications can be distributed broadly to users to fulfill their daily usages, therefore it may be easier to collect text entry specific corpus [6] to meet the requirement of contextual inquiry.

It is also interesting to us if ACC and suggested context length are useful to alphabetic language text entry such as T9, LetterWise or WordWise. For those mobile device oriented text input methods, however, the restricted memory space is an unavoidable issue. How to walk in the fine line between space complexity and AAC is one of our next lessons to learn. By not considering the learning cost of a new text entry device, we hope to completely define the amortized modification cost, such that one text entry evaluation approach can be used for all input methods, devices, and languages.

**APPENDIX: CONTRIBUTION AND BENEFITS**
An automated Chinese text entry evaluation combines cursor movement and candidate selection time into the amortized cost as a novel metric, and suggests appropriate context lengths at the minimum cost.